\documentclass[12pt,preprint]{aastex}

\def \eg{{e.g.,}}
\def \etal{{et~al.\null}}
\def \ie{{i.e.,}}

\slugcomment{}

\shorttitle{Resolved Stellar Population of a Virgo dSph}
\shortauthors{Durrell et al.}

\begin{document}

\title{The Resolved Stellar Populations of a Dwarf Spheroidal Galaxy in the Virgo Cluster}

\author{Patrick R. Durrell\altaffilmark{1}, Benjamin F. Williams\altaffilmark{2}, Robin Ciardullo\altaffilmark{2}, John
J. Feldmeier\altaffilmark{1,11}, Ted von Hippel\altaffilmark{3}, Steinn
Sigurdsson\altaffilmark{2}, George H. Jacoby\altaffilmark{4}, Henry
C. Ferguson\altaffilmark{5}, Nial R. Tanvir\altaffilmark{6}, Magda
Arnaboldi\altaffilmark{7}, Ortwin Gerhard\altaffilmark{8}, J. Alfonso 
L. Aguerri\altaffilmark{9}, Ken Freeman\altaffilmark{10}, and Matt
Vinciguerra\altaffilmark{2}}

\altaffiltext{1}{Department of Physics \& Astronomy, Youngstown State 
University, Youngstown, OH 44555; prdurrell@ysu.edu; jjfeldmeier@ysu.edu}
\altaffiltext{2}{Department of Astronomy \& Astrophysics, 
Pennsylvania State University, University Park, PA 16802; 
bwilliams@astro.psu.edu; rbc@astro.psu.edu; steinn@astro.psu.edu}
\altaffiltext{3}{The University of Texas, Department of Astronomy, 
1 University Station C1400, Austin, Texas 78712; ted@astro.as.utexas.edu}
\altaffiltext{4}{WIYN Observatory, 950 North Cherry Avenue, P.O. Box 26732, 
Tucson, AZ 85726; jacoby@wiyn.org}
\altaffiltext{5}{Space Telescope Science Institute, 3700 San Martin Drive, 
Baltimore, MD 21218; ferguson@stsci.edu}
\altaffiltext{6}{Department of Physics \& Astronomy, University of Leicester, Leicester LE1 7RH, UK; 
nrt3@star.le.ac.uk}
\altaffiltext{7}{European Southern Observatory, Karl-Schwarzchild-Str.~2,
85748 Garching, Germany, marnabol@eso.org}
\altaffiltext{8}{Max-Planck-Institut fuer Extraterrestrische Physik,
P.O. Box 1312, D-85741 Garching, Germany, gerhard@exgal.mpe.mpg.de}
\altaffiltext{9}{Instituto de Astrofisica de Canarias, C/ V\'{i}a L\'{a}ctea, 
s/n, 38205, La Laguna, Tenerife, Spain; jalfonso@ll.iac.es}
\altaffiltext{10}{Mount Stromlo Observatory, Research School of Astronomy 
and Astrophysics, Mount Stromlo Observatory, The Australian National 
University, ACT 0200 Australia; kcf@mso.anu.edu.au}
\altaffiltext{11}{NSF Astronomy \& Astrophysics Postdoctoral Fellow}
\begin{abstract}

We report on the discovery of a faint ($M_V \sim -10.6 \pm 0.2$) dwarf 
spheroidal galaxy on deep F606W and F814W \textsl{Hubble Space
Telescope} images of a Virgo intracluster field.  The galaxy is easily
resolved in our images, as our color magnitude diagram (CMD) extends
$\gtrsim 1$~magnitude beyond the tip of the red giant branch (RGB).  Thus, it
is the deepest CMD for a small dwarf galaxy inside a cluster environment.
Using the colors of the RGB stars, we derive a metal abundance for the dwarf 
of [M/H]$= -2.3 \pm 0.3$, and show that the metallicity dispersion is 
less than 0.6~dex at 95\% confidence.  We also use the galaxy's lack of AGB 
stars and the absence of objects brighter than $M_{bol} \sim -4.1 \pm 0.2$ 
to show that the system is old ($t \gtrsim 10$~Gyr).  Finally, we derive the 
object's structural parameters, and show that the galaxy displays no obvious 
evidence of tidal threshing.   Since the tip of the red giant branch distance 
($(m-M)_0 = 31.23 \pm 0.17$ or $D = 17.6 \pm 1.4$~Mpc) puts the galaxy 
near the core of the Virgo cluster, one might expect the object to have undergone
some tidal processing.  Yet the chemical and morphological similarity 
between the dwarf and the dSph galaxies of the Local and M81 Group 
demonstrates that the object is indeed pristine, 
and not the shredded remains of a much larger galaxy.  We discuss the 
possible origins of this galaxy, and suggest that it is just now falling into 
Virgo for the first time.

\end{abstract}


\section{Introduction}

An understanding of the processes that affect the formation,
evolution, and sometimes destruction of dwarf galaxies is critical to
our overall picture of galaxy formation.  Since dwarf galaxies are the
most common type of galaxy in the Universe, any model of galaxy
formation is incomplete if we cannot understand these objects.
Moreover, under the current paradigm of hierarchical structure
formation \citep[\eg][]{kauf93, cole00}, massive galaxies are formed
from the mergers of less massive objects.  Thus, dwarf galaxies are an
important building block of the universe, and an understanding of
their properties will help illuminate the processes of galaxy
formation more generally.

The faintest examples of dwarf galaxies are the dwarf spheroidals
(dSphs).  Using the nomenclature of \citet{ggh03}, these are
low-luminosity ($M_V \gtrsim -14$), low central surface brightness ($\mu_V
\gtrsim 22$~mag arcsec$^{-2}$) objects with shallow radial profiles
and stellar populations that are dominated by old and
intermediate-aged stars\footnote{dSph galaxies are the faint end of
the dwarf elliptical [dE] galaxy sequence, and when we note dE
galaxies we are including those gas-poor dwarf galaxies more luminous
than $M_V=-14$}.  Most of what we know about dSphs comes from
observations of $\sim 20$~galaxies in the low-density environment of
the Local Group.  From the data, it appears that dSphs are fairly
homogeneous in nature, with small, but dense dark matter halos and low
metallicities that are in proportion to their luminosity \citep[for a
review see][]{mateo98}.

Much less is known about dSphs in the cluster environment.
Surveys in Virgo and Fornax have successfully found large numbers of dE
and bright dSph galaxies \citep{vcc, fcc, phil98, th02, sab03}, but there is
still considerable uncertainty in just how many of these lowest-luminosity
objects exist \citep[\eg][]{hmi03}.  The scarce information we do have on 
cluster dSph galaxies suggests that their properties are similar to those 
of Local Group dwarfs \citep{hmi05, cald06}.  If correct, then such a result 
is somewhat surprising, since the tidal forces of the cluster environment are
expected to take their toll on these objects.  Indeed, \citet{bekki01}
have shown that dSphs can be stripped of their outer stars in
only a few Gyr.  This would imply that both the morphology and
chemistry of cluster dSphs should differ significantly from their
Local Group counterparts.

Until recently, abundance studies of cluster dSph galaxies (based on
photometry of individual stars) have been impractical.  For example,
although \citet{har98} were able to resolve the brightest red giant
branch (RGB) stars in a Virgo cluster dE,N galaxy using a 32~ks exposure
with the \textsl{Hubble Space Telescope's WFPC2} camera, these data were
limited to a single filter (F814W).  However, with the advent of the 
\textsl{Advanced Camera for Surveys (ACS),} it is now possible to measure the 
broad-band colors of red giants in these systems.  Indeed, 
\citet{cald06} has recently used the \textsl{ACS} to resolve the stellar 
populations of four dE/dSph galaxies in the Virgo cluster, and obtain 
color-magnitude diagrams of their stars down to $I\sim 28$, making it
the deepest study of individual stars in cluster dwarf galaxies.  Their data 
appear to confirm that the dE/dSph galaxies of Virgo obey the same
luminosity-metallicity relation as the dE/dSph galaxies of the Local Group.

Here, we present an extremely deep F606W and F814W study of a faint
($M_V \sim -10.6$) dSph that was serendipitously found during a survey of 
intracluster stars in Virgo \citep{VICS1}.  We use the data, which reach 
$I \sim 28.4$, to show that the galaxy is remarkably similar to undisturbed 
dSphs in the Local Group, both in morphology, and in stellar content.
In Section~2, we describe both our data reductions and our artificial star
experiments, and present a color-magnitude diagram of the dSph's 
red giant branch stars.  In Section~3, we use these data to derive the
system's distance as well as its stellar population.  We show that the 
galaxy has a mean metallicity that is very low ([M/H] $\sim -2.3 \pm 0.3$),
and is composed entirely of stars older than 8 to 10~Gyr.  We also
derive the galaxy's structural parameters, and show that its central
surface brightness and core radius are typical of dSphs in the Local
Group.  In Section~4, we compare the galaxy's properties to those
of Local Group objects, and attempt to investigate the effects of
environment on its history.  We show that the galaxy displays no obvious 
evidence of tidal disruption, and has a mean metallicity appropriate 
for its luminosity.   We conclude by considering the possible
origin of this object.

\section{Observations and Data Analysis}

Between 30 May 2005 and 7 June 2005 we used the \textsl{Advanced Camera
for Surveys} on the \textsl{Hubble Space Telescope} to obtain deep
F606W and F814W images of a Virgo intracluster field
($\alpha(2000) = $12:28:10.80, $\delta(2000) = $12:33:20.0,
orientation 112.58 degrees).  This field, which is projected $\sim 180$~kpc
from M87 \citep[assuming a mean Virgo distance $d= 16$ Mpc; \eg][]{jbb04} was 
chosen to be near the cluster center, but away from any known object:  the
closest cataloged galaxy is the dwarf elliptical VCC~1051, which is 
$\sim 5\arcmin$ (24~kpc) to the northwest; the nearest large galaxy, M86,
is over 170~kpc away.  

Our F814W ($I$-band) data consisted of 22 exposures, with 26880~s of 
integration time; the F606W (wide $V$-band) observation included 52
exposures totaling 63440~s.  These data were co-added, and re-sampled 
using the {\tt multidrizzle}\footnote{multidrizzle is a product of the Space
Telescope Science Institute, which is operated by AURA for NASA.
http://stsdas.stsci.edu/pydrizzle/multidrizzle} task within 
PyRAF\footnote{PyRAF is a product of the Space Telescope Science Institute, 
which is operated by AURA for NASA.} \citep{koekemoer} to
produce images with $0\farcs 03$~pixel$^{-1}$.  The details of these
reductions, and an image of the field illustrating its position
in the cluster is given by \citet{VICS1}.

\begin{figure}
\epsscale{1.00}
\plotone{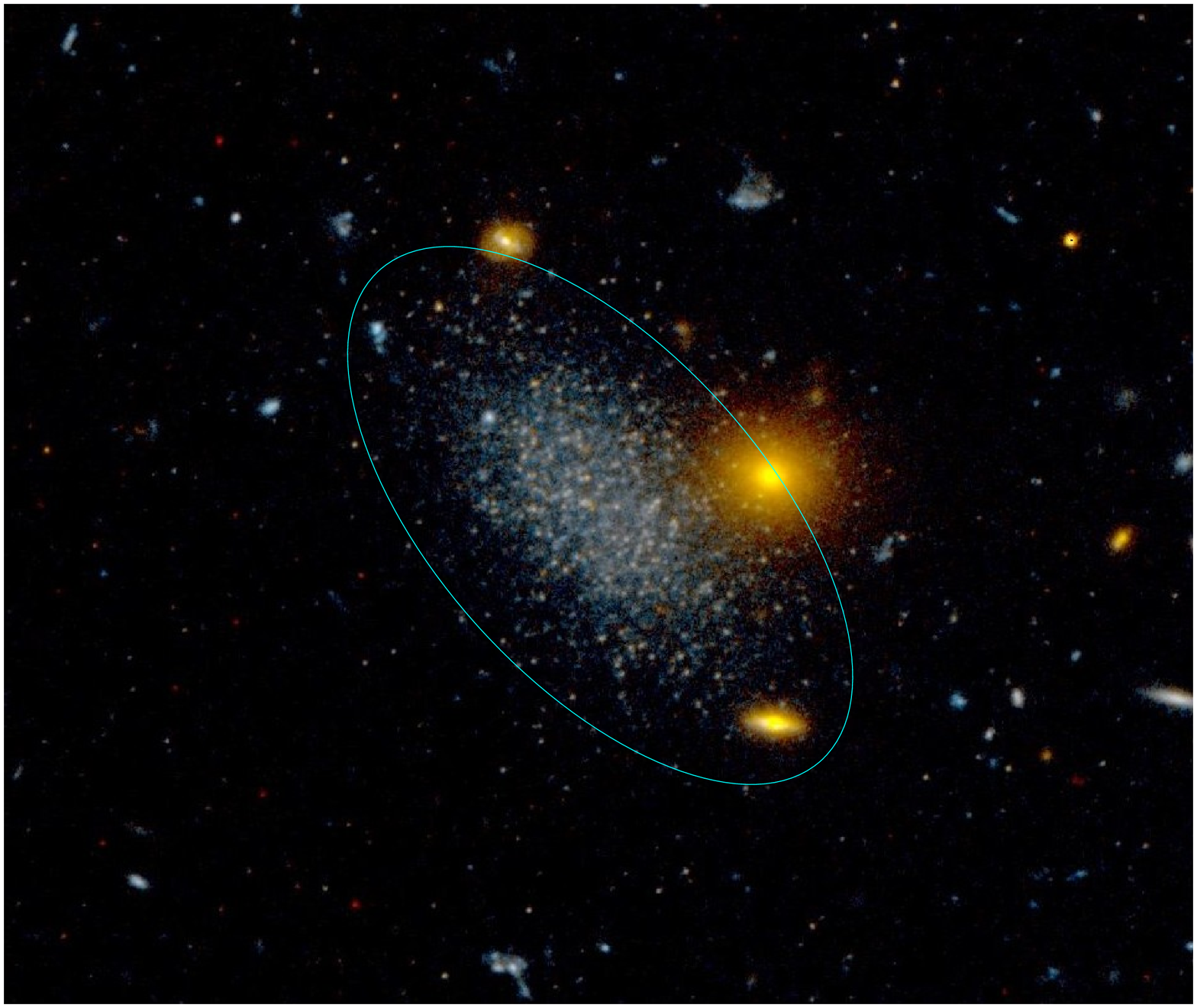}
\caption{Color image of a $28\arcsec \times 24\arcsec$ section of our
image, centered on the dSph galaxy.  In the image, blue represents
$2 \times$ F606W$-$F814W, green represents F606W, and red represents
F814W\null.  The ellipse denotes the boundary used to defined a
subsample of stars which minimizes contamination (see text).
North is to the top, and East is to the left.
\label{image}}
\end{figure}

Immediately after data acquisition, an inspection of our images revealed
the presence of a previously unknown galaxy in the field.  The object,
whose center is located at $\alpha(2000) = $12:28:15.5, $\delta(2000) = 
$+12:33:37.0 (uncertainty $\sim 0\farcs 2$) is $\sim 15\arcsec$ in extent
and clearly resolved into stars.  A color image of the object is shown
in Figure~\ref{image}.
 
Our point source photometry was similar (but not identical to) that
performed by \citet{VICS1}.  To avoid dealing with variations in the
\textsl{ACS} point spread function (PSF), we began by limiting our
photometric reductions to a $2200 \times 1600$ pixel ($66\arcsec\times
48\arcsec$) region surrounding the galaxy.  We then chose three
bright, isolated stars on our F606W image, and four bright isolated
stars on the F814W image to define the PSFs, and used DAOPHOT~II and
ALLSTAR to perform the photometry \citep{stet87, stet92}.  Two
DAOPHOT~II/ALLSTAR passes were performed to detect as many of the
stars as possible.  The F606W and F814W datasets were then merged with
a 1.5 pixel matching criterion to create a preliminary object catalog
of $\sim 1000$~objects.  Finally, we used the DAOPHOT $\chi$ and
$r_{-2}$ image parameters to remove background galaxies and severely
blended stars from the list \citep{stet87, kron80}.

Our instrumental magnitudes were placed on the VEGAMAG system using the 
prescription and zero points provided by \citet{sir05}.  To obtain the
offset between our ALLSTAR magnitudes and the VEGAMAG system,
we used the profiles of the PSF stars on each image;
the rms scatter in this calibration was $\sim 0.02$~mag.   We note that
since the region under consideration is small (only $\sim 8\%$ of the
entire \textsl{ACS} field), differential effects associated with charge
transfer efficiency were negligible ($\sim 0.01-0.02$~mag), and were
ignored.

\subsection{Artificial Star Experiments}

To ascertain the photometric uncertainties and incompleteness in our
data, we used the traditional method of adding simulated stars of
known brightness to the science frames and re-reducing them following
the exact same procedures as for the original data.  To each frame, we
added 9000 stars (300 runs with 30 stars added per run) to an
elliptical region centered on the dwarf galaxy, and re-executed our
two-pass DAOPHOT II/ALLSTAR photometric procedure, including the
rejection of non-stellar sources, and the merging of the datasets.  By
limiting the number of fake stars to 30 per run, we did not
significantly alter the crowding conditions on the images; by defining
the mean F606W$-$F814W color of our artificial stars to be roughly
equal to that of the observed objects (F606W$-$F814W = 1.0), we
ensured that our experiments adequately reproduced losses during the
catalog merging and image classification processes.  Finally, by
defining the stars' luminosity function to be a rising exponential
between F814W = 22 and F814W = 29, we improved our statistics at the
faint end of the luminosity function, where the uncertainties are
largest.  The result of these experiments were the photometric
completeness function $f(m)$, defined as the ratio of the stars
recovered to stars added {\it regardless of the magnitude that was
actually measured,} $\Delta(m)$, the mean shift in magnitude of the
stars ($\Delta(m) = m_{out} - m_{in}$), and $\sigma(m)$, the
dispersion about this mean.  All of these functions were computed in
0.5~mag bins, to ease the computational requirements.

\begin{figure}
\epsscale{1.00}
\plotone{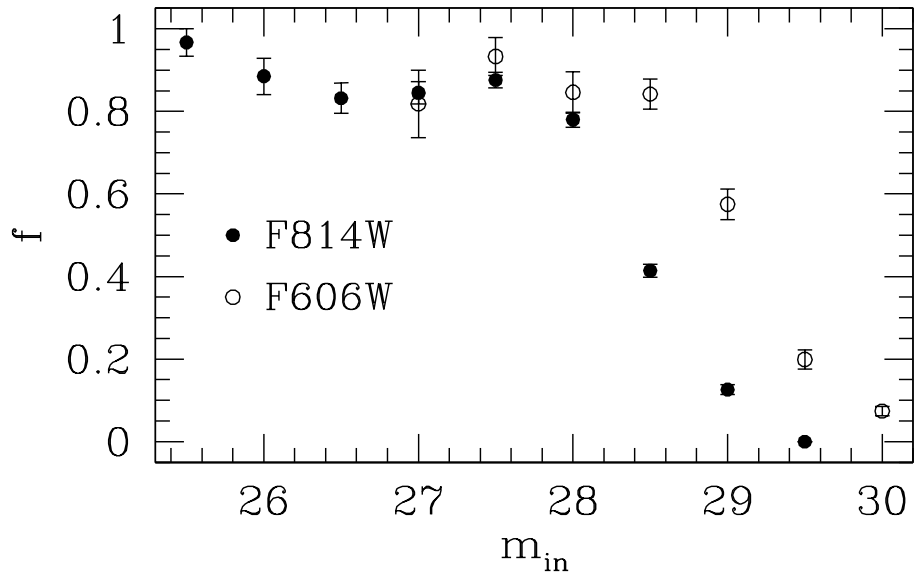}
\caption{Fraction $f$ of recovered artificial stars for the F814W (filled circles) 
and the F606W (open circles) images.   The F606W data is based on added stars with input colors 
F606W$-$F814W=2.0, and that of the F814W data is from stars with input colors 
F606W$-$F814W=0.0.
\label{completeness}}
\end{figure}

To ascertain each filter's limiting magnitude (defined as the 50\% completeness
fraction), we performed similar artificial star tests using a variety of 
F606W$-$F814W input colors.    The results from these experiments are illustrated 
in Figure~\ref{completeness}.     From these experiments, we obtained
$m_{lim} = 29.1 \pm 0.1$ in F606W and $28.4 \pm 0.1$ in F814W, where the errors
reflect both the rms scatter due to the use of different colors, and the
uncertainty associated with varying levels of crowding within the galaxy.
As expected, these limits are slightly brighter than those found by 
\citet{VICS1}, whose analysis focused on the uncrowded regions of the frame.
Moreover, our classification criteria kept only those objects with the
best photometric accuracy: obvious galaxies and extremely blended
objects were removed from the catalog.  As result, even at F814W = 27,
$\sim 10\%$ of the artificial stars were not recovered.  Fortunately,
since we are not making any inferences based solely on the absolute
number of stars detected, these missing objects do not affect our
analyses in any significant way.

\begin{figure}
\epsscale{1.00}
\plotone{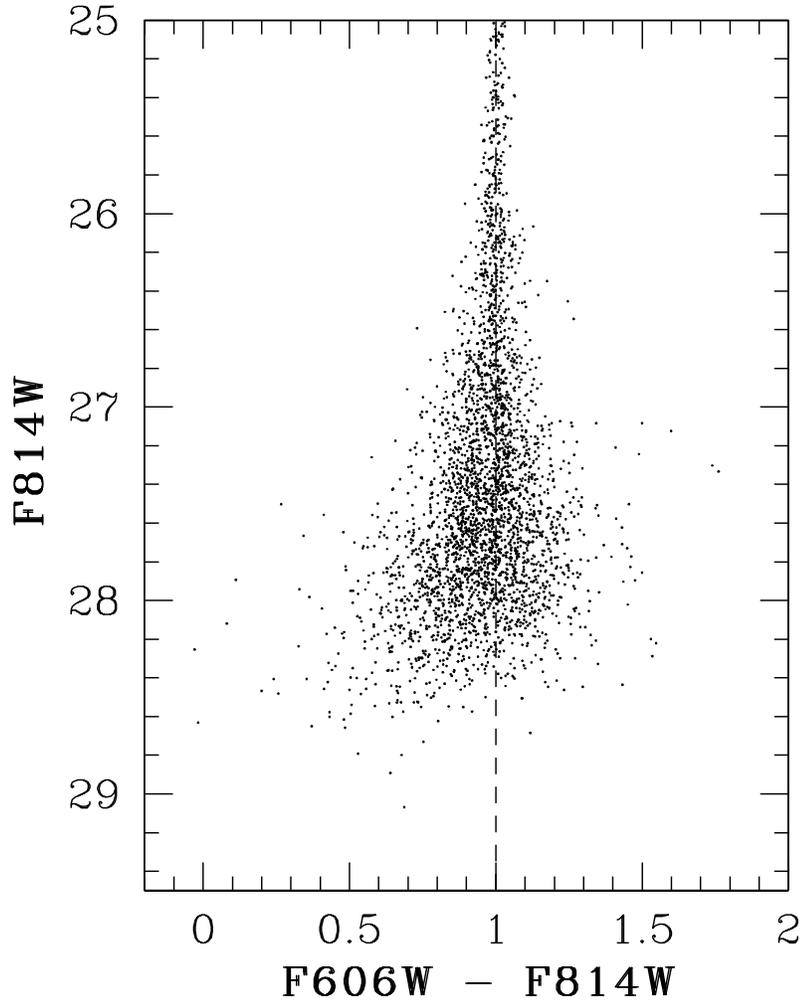}
\caption{Output color-magnitude diagram for our artificial stars. 
The dashed line denotes the input colors for the stars (F606W$-$F814W)=1.00. 
Note the slight offset in the colors at faint magnitudes.  This is
due to the different limiting magnitudes of the two filters.
\label{fakestars}}
\end{figure}

A color-magnitude diagram for our artificial stars is shown in
Figure~\ref{fakestars}.  The figure illustrates that the photometric errors
increase with magnitude, so that by $m_{lim}$, the uncertainties are
$\sim 0.2$~mag.  The figure also shows that there is a small blueward shift
at faintest levels;  this is due to the uneven depth of the two images.
(At the faintest levels, our F814W photometry only detects objects
whose photometric errors are in the positive direction.)  Fortunately, 
down to the limiting magnitudes of the frames, this color shift is $<
0.1$~mag.  These corrections, while small, will be accounted for in
the following sections.

\subsection{Color Magnitude Diagram}

\begin{figure}
\plotone{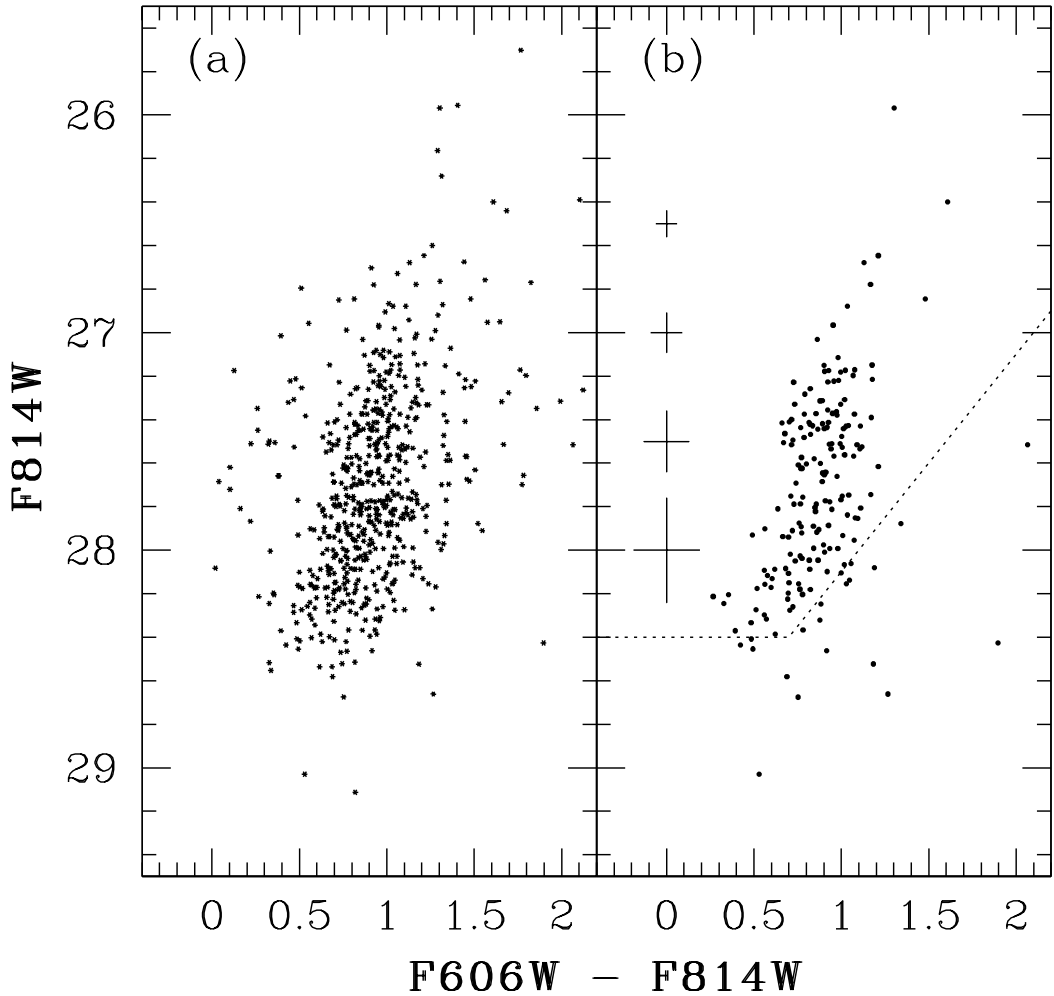}
\caption{(a) The color-magnitude diagram (in the VEGAMAG system) for the 
611 stellar objects located in a $66\arcsec \times 48\arcsec$ region 
centered on the dSph galaxy.   (b) The `dwarf-only' CMD, formed from 
a subset of 181 stars located within the inner elliptical region shown in 
Figure~\ref{image}.  The dotted lines denote the
50\% completeness levels, while the error bars represent the typical 
photometric uncertainties.  Note the discontinuity at F814W $\sim 27.1$;
this is the tip of the red giant branch.
\label{cmd}}
\end{figure}

The color-magnitude diagram for all the stellar objects in our field
is displayed in the left-hand panel of Figure~\ref{cmd}.  The diagram
shows that most objects lie in a sequence near F606W$-$F814W $\sim 1$
that extends up to F814W $\sim 27.1$.  However, there is a significant
amount of scatter about this line.  This scatter comes from two
sources: the RGB stars that pervade Virgo's intracluster space
\citep{ftv, dur02, VICS1}, and, to a lesser extent, unresolved
background galaxies.  To minimize the effects of these contaminants,
we identified a subsample of stars within the galaxy's F814W $\sim
26.5$~mag~arcsec$^{-2}$ isophote.  This subsample, which is defined
via an ellipse with semi-major axis $8\farcs 1$, an axis ratio of $b/a
= 0.5$, and a position angle of $43^\circ$ (see Section 3.4) should
have minimal contamination: based on star counts in the rest of our
field, only $\sim 16$ out of the 171 objects brighter than $F814W=28.4$
should be contaminants.  Thus, we can use the data from this
103~arcsec$^2$ region to form a ``clean'' color-magnitude diagram
of the galaxy's stars.  The elliptical region which defines our
subsample is shown in Figure~\ref{image}; the right-hand panel of
Figure~\ref{cmd} shows its CMD.

\section{Analysis}

The CMD in Figure~\ref{cmd} displays the hallmarks of a red-giant branch
population with F814W $> 27$.  There are few stars brighter than this 
limit, and no evidence of any bright blue stars arising from a young
stellar population.  Since the data clearly extend more than 1~mag down
the RGB, we can use the diagram to investigate both the distance and
metallicity of the galaxy.

\subsection{TRGB Distance}

The tip of the red giant branch (TRGB) in the $I$-band has repeatedly
been shown to be an excellent distance indicator for old, metal-poor
populations.  In stellar systems with [M/H] $< -1$, the TRGB is
essentially independent of both age and metallicity
\citep[\eg][]{lfm93, sakai96, fer00, bell01, mcc04, mou05, cald06}, so
all that is required is a measure of the bright-end truncation of the
RGB and an estimate of foreground reddening
\citep[$E(B-V) = 0.025$;][]{schlegel98}.  Our only limitation is the
small size of the galaxy, which restricts the number of stars available for
analysis.

Although foreground/background contamination in our `dwarf-only' CMD
(right panel of Figure~\ref{cmd}) is small, to make our TRGB detection
as unambiguous as possible we have further restricted our sample to
objects with F606W$-$F814W $< 1.3$.  Stars redder than this are not on
the metal-poor RGB and are unlikely to be members of the galaxy.  This
makes very little difference to the analysis, since it excludes only 3
objects from consideration.  Moreover, since the RGB tip for red,
metal-rich stars is fainter than that for blue objects, eliminating
these stars should not affect our distance determination.  We note
that the \textsl{ACS} F814W filter bandpass is very similar to that of
the traditional $I$-band: according to
\citet{sir05}, the transformation between these two systems is
$I = F814W - 0.008$ (for stars with F606W$-$F814W=1) in the VEGAMAG system. 

Visual inspection of the Figure~\ref{cmd} shows a rather sharp transition at
F814W $\sim I \sim 27.1$, albeit with a small number of stars.  While
the increased Poissonian noise from small numbers can lead to a possible
systematic bias in $I_{TRGB}$, \citet{mf95} have shown that with at
least $\sim 100$ stars in the top magnitude of the RGB, such biases are
small.  Since our analysis includes $\sim 130$~objects, we are safely
above this threshold. 

Another possible source of systematic error is image crowding:
in dense fields, stellar blends can make the RGB tip appear brighter
than it should be.  However, in our case, this effect should not
be important.  Our artificial star experiments demonstrate that the
mean magnitude shift at F814W $\sim 27$ is very small, less than
0.01~mag.  This result, which is confirmed by additional artificial
star experiments with F814W $< 27.1$, is due principally to our
image classification criteria, which rejects the most heavily
blended objects (\ie\ those most likely to bias the results towards brighter 
magnitudes).

A quantitative estimate of the RGB tip location can be derived using
the Sobel edge-detection filter of \citet{lfm93}.  This technique, which
was originally used on binned functions, has since been modified
to work on discrete stellar samples by treating each object as a 
Gaussian distribution with a dispersion equal to the expected photometric 
error \citep{sakai96}.    We note that more rigorous techniques have been 
employed to derive $I_{TRGB}$, such as the method of maximum-likelihood 
\citep{mendez02, mou05}, and the second derivative of the
luminosity function \citep{cioni00}.  However, maximum-likelihood
methods are better for datasets where the RGB power-law slope is observed 
for over a magnitude.  In our case, photometric incompleteness and image 
crowding make the use of the technique problematic.  Moreover, the small 
number of stars limits the reliability of second derivative calculations.
Because our background contamination is so small, the simpler Sobel
edge-detection filter should be  adequate for our purpose.  

\begin{figure}
\plotone{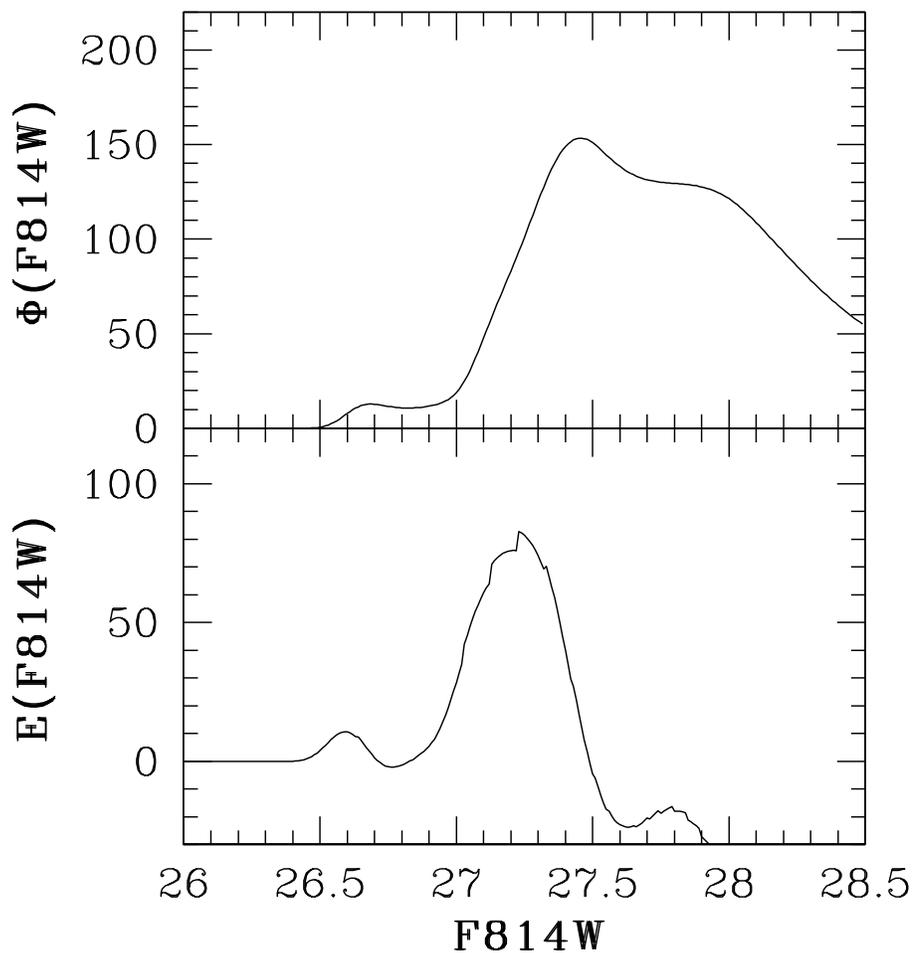}
\caption{The top panel shows a continuous luminosity function formed 
by co-adding the Gaussian representation of stars in our minimally 
contaminated subsample (see text).  Objects with F606W$-$F814W $> 1.3$ have 
been excluded from the analysis.   The bottom panel is the result of
applying the Sobel edge-detection algorithm.  This panel suggests that
the tip of the red giant branch is at F814W$_{TRGB} = 27.22 \pm 0.15$.
\label{sobel}}
\end{figure}

To employ the Sobel-edge detection method, we treated each star as a
Gaussian distribution with a mean equal to its observed magnitude (plus the 
small $\Delta m$ shift predicted by our artificial star experiments), and
a dispersion, $\sigma(m)$, equal to the photometric uncertainty at that 
magnitude.  We then co-added the Gaussians, to produce the luminosity
function shown in the top panel of Figure~\ref{sobel}.  Even with the
large photometric errors ($\sigma \sim 0.07$ at F814W = 27), the sharp 
rise of the luminosity function is evident, as is a small
contribution from brighter stars, which presumably arise from the
galaxy's AGB component (more on this below).

To determine the location of the tip of the red giant branch, we applied
to our continuous luminosity function the edge-detection algorithm
\begin{equation}
E(m) = \Phi (m+ \sigma(m)) - \Phi(m - \sigma(m))
\end{equation}
where $\sigma(m)$ is the photometric error defined via our artificial
star experiments \citep{sakai96}.  This function, which is displayed
in the bottom panel of Figure~\ref{sobel}, reveals a rather wide peak
near F814W $\sim 27.2$.  From the figure, the location of the RGB
discontinuity is at $I_{TRGB} = 27.22 \pm 0.15$, where the uncertainty
is based on the full-width-half-maximum of the distribution.  
This error is likely conservative \citep[\eg][]{sakai96}, though it is
similar to that expected based on the number of stars available
\citep{mf95}.

To derive the distance to the dwarf, we adopt $M_{I,TRGB}=-4.06 \pm
0.07$ (random error) as the absolute magnitude of the metal-poor RGB
tip \citep{fer00}.  This number is very similar to that of other
recent determinations: the empirical calibration of \citet{da90}
yields $M_{I,TRGB}= -3.96 \pm 0.06$ for objects with [Fe/H]$=-2.3$,
while a more recent determination by \citet{bell01} gives $M_{I,TRGB}
= -4.04 \pm 0.12$ for [Fe/H]$=-1.7$. (The metallicity dependence of
$M_{I,TRGB}$ over this range is much smaller than these random
errors.)  Thus we derive $(m-M)_I=31.28 \pm 0.17$ for the dwarf.
Assuming a foreground reddening of $E(B-V) = 0.025$ \citep{schlegel98}
\citep[and thus $A_{F814W} \sim A_I = 0.046$; ][]{sir05}, we derive a
distance modulus of $(m-M)_0 = 31.23 \pm 0.17$, or $D = 17.6\pm
1.4$~Mpc (random error only).   This distance is similar to the
Cepheid, surface brightness fluctuation, and planetary nebula
luminosity function distances to the core of Virgo \citep{jacoby90,
freedman01, tonry01}.  Although the three-dimensional structure of the
Virgo cluster is complex, the line-of-sight depth of the Virgo core
(as defined by the early-type galaxies) surrounding M87 is at least
$\sim 2-3$ Mpc \citep[\eg][]{nt2000, jbb04}, and the distance to the 
object is consistent with being located near the center of the
cluster.

\subsection{Metallicity}

As noted earlier, much of the color spread in the dSph's RGB is
due to photometric errors.  Thus, to measure metallicity, our approach was 
to derive a {\it mean\/} abundance for the galaxy, and then attribute any
additional scatter on the RGB to a dispersion in metallicity.  
Since the preceding analysis has shown that the galaxy contains
no significant AGB population (also see Section 3.4), we 
can ignore the alternative that the color spread is due to the
presence of young or intermediate-age stars.

\begin{figure}
\plotone{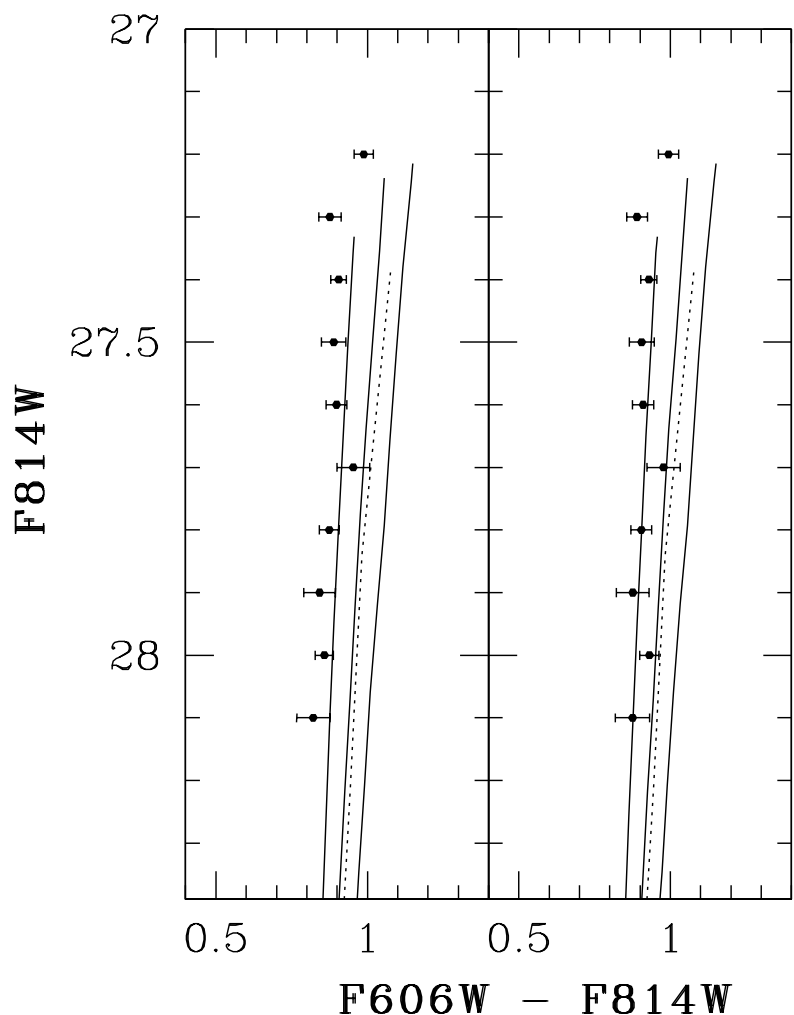}
\caption{A comparison of observed red giant branch of our dSph galaxy
with the 12.5~Gyr models of \citet{gir05}.  The solids are (from left to
right) $Z=0.0001, 0.0004$, and 0.001.  The dashed lines show the
M92 fiducial ($Z\sim 0.0003$) from \citet{brown05}.  All the curves have 
been shifted to a distance modulus of $(m-M)_{F814W}=31.29$  and reddened by
$E({\rm F606W}-{\rm F814W}) = 0.025$.  The points represent the observed
mean red giant branch, while the error bars show the uncertainties in the 
mean.  The left panel displays the raw measurements; the right panel shifts
the observed colors by an amount predicted by our artificial star 
experiments.  
\label{ridgeline}}
\end{figure}

To derive the mean metallicity of the dSph galaxy, we compared the
ridge line derived from the CMD in Figure~\ref{cmd} with the
scaled-solar abundance isochrones of \citet{gir05}, which are the Padova isochrones
\citep{gir00, gir02} transformed directly to the \textsl{ACS/WFC}
filter system.  In addition, we have also used the observed fiducial
sequences from \citet{brown05}, which are derived from photometry of
Milky Way clusters covering a wide range in metallicity.
Figure~\ref{ridgeline} compares the \citet{gir05} 12.5~Gyr sequence
(shifted by $(m-M)_{\rm F814W}= 31.29$ and $E({\rm F606W}-{\rm
F814W})=0.025$) to the observed colors of the galaxy's RGB stars, both
with and without the $\Delta m$ magnitude shifts predicted by our
artificial star experiments.  Also shown for comparison is the M92
fiducial sequence of \citet{brown05}.  According to \citet{brown05},
this cluster has an [Fe/H] value of $-2.14$, but is enhanced in
$\alpha$-process elements by [$\alpha$/Fe] $= +0.3$, thus implying $Z
\sim 0.0003$.  Consequently, its position in the color-magnitude
diagram is consistent with the $Z=0.0004$ scaled-solar Padova
isochrone.

As Figure~\ref{ridgeline} illustrates, the dSph galaxy is extremely
metal-poor: the colors of its RGB stars are bluer than the metal-poor
M92 fiducial and very near the $Z=0.0001$ Padova isochrone.  For a
more quantitative estimate, we can use the mean color of the RGB at
F814W $= 27.6\pm 0.1$ (or $M_{\rm F814W} = -3.7$ at the distance of
the dwarf) and interpolate in the Padova models; this particular
magnitude was chosen to be deep enough to adequately sample the RGB
(\eg not at the RGB tip), but not so faint as to be adversely affected
by large photometric errors and photometric incompleteness.  This
procedure yields a mean metallicity for the system of [M/H] $= -2.3
\pm 0.3$, where the quoted error is that of the mean, not of the
distribution.  This error does not include possible systematic errors
in the Padova models, but does include a $\pm 0.1$~dex uncertainty due
to distance.  The errors in our color calibration ($\sim 0.03$~mag)
and the reddening ($\sim 0.02~$mag) contribute an additional $\pm
0.3$~dex uncertainty to our determination.

To estimate an upper limit to the metallicity spread in the galaxy, we 
compared the observed width of the RGB to the expected
dispersion derived from our artificial star experiments.  To do this,
we used the data displayed in the right-hand panel of Figure~\ref{cmd},
while excluding the one, extremely red (F606W$-$F814W = 2.06, F814W = 27.5)
object from consideration.   We then formed the sum
\begin{equation}
\chi^2 = \sum_i { \left( c_{obs} - c_{iso} \right)^2 \over \sigma_c^2 }
\end{equation}
where $c_{obs}$ is the observed F606W$-$F814W color of each star with
$27.3 < {\rm F814W} < 28.0$, $c_{iso}$, the predicted color for the
star (given its observed F814W magnitude and a system metallicity of
[M/H] $= -2.3$), and $\sigma_c$ the expected color dispersion at
magnitude F814W, derived from our artificial star experiments.  The
resulting value, $\chi^2/\nu = 1.16$ for 95 degrees of freedom,
indicates that the true red giant branch is slightly wider, but
not inconsistent with, the distribution expected from a single
metallicity population.  This same type of analysis also demonstrates
that any additional (intrinsic) color dispersion on the red giant
branch must be less than $\sigma_{{\rm F606W}-{\rm F814W}} \sim 0.09$
at the 95\% confidence level.  This limits the metallicity spread of
our dSph galaxy to $\lesssim 0.6$~dex.

\subsection{AGB Stars and the Age of the Stellar Population}

The presence of stars brighter than the RGB tip have often been
used to ascertain the existence of intermediate-age ($t < 10$~Gyr)
stars: the more luminous the tip of the AGB, the younger the stellar
population \citep[\eg][]{ma82, arm93, cald98}.   Our CMD shows 
that our dwarf spheroidal galaxy contains very few stars brighter than F814W 
$\sim 27.1$.  However, since the number of stars present in the galaxy is
small, it is difficult to make precise measurements of the AGB phase of
evolution.  Moreover, as \citet{VICS1} have demonstrated, our dwarf 
galaxy resides within a sea of intracluster stars, so any RGB stars 
foreground to the dwarf can masquerade as `brighter' AGB objects.

Nevertheless, it is possible to place a limit on the presence (or absence) of
AGB stars in our galaxy.  To do this, we used our derived distance to the 
dSph and the stars' observed colors to estimate the bolometric magnitudes 
of the galaxy's brightest objects.  Specifically, we converted the stars'
F814W$-$F606W colors to $(V-I)$ using the color terms of \citet{sir05},
and applied the \citet{da90} bolometric correction
\begin{equation}
BC_I = 0.881 - 0.243 (V-I)_0
\end{equation}
to generate values of $M_{bol}$ for all objects fainter than
F814W $= 26.0$.  This procedure is uncertain at the $\sim 0.2$~mag
level:  in addition to $\sim 0.05$~mag errors associated with the
color transforms and bolometric correction, our distance to the
galaxy is uncertain by $\sim 0.15$~mag.  Nevertheless, this
calculation is sufficiently accurate for our purpose.

\begin{figure}
\plotone{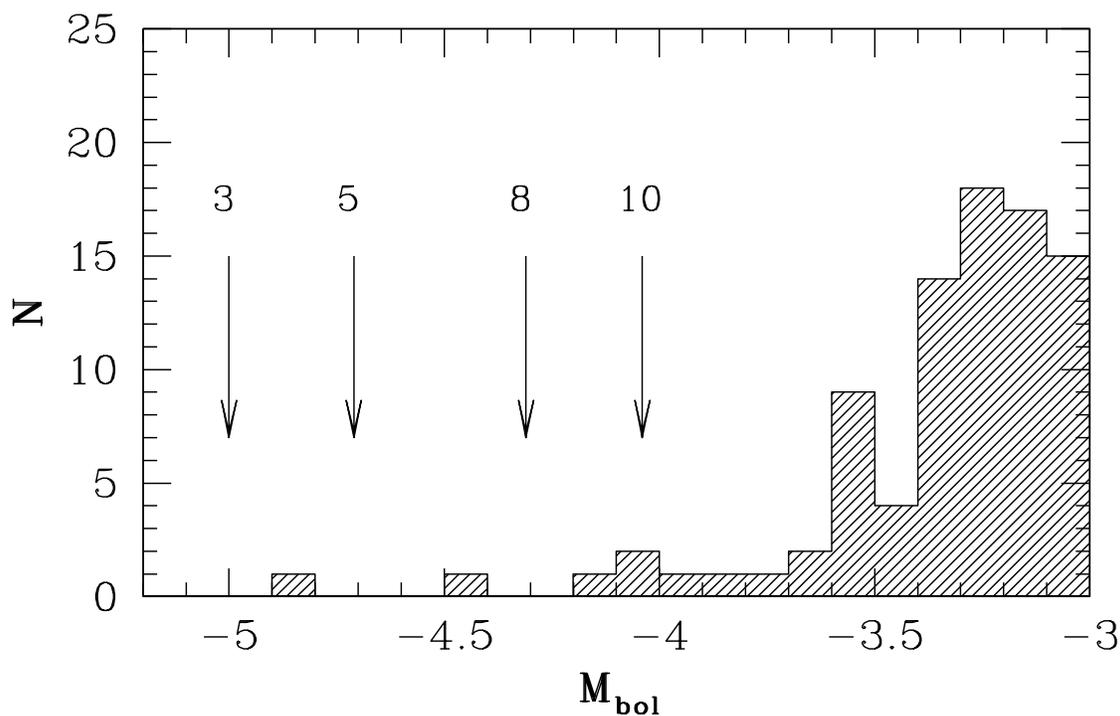}
\caption{A histogram of bolometric magnitudes for the brightest stars in
our minimally contaminated dSph subsample; no corrections for
photometric incompleteness have been applied.  The arrows denote the
maximum attainable brightnesses of AGB stars formed from populations
with ages of 3, 5, 8, and 10~Gyr \citep[from Figure~19 of][]{rej06}.
The two brightest stars are likely interlopers.  The data suggest that
the dSph population is at least 8~Gyr old.
\label{agb}}
\end{figure}

A histogram of the derived values of $M_{bol}$ for the brightest stars
in the right-hand panel of Figure~\ref{cmd} is shown in
Figure~\ref{agb}.  Also plotted are the maximum attainable
luminosities for AGB stars produced by 3, 5, 8, and 10~Gyr old
populations \citep{rej06}.  As the figure illustrates, our sample of
stars contains two objects that may indicate the existence of a
(small) number of intermediate age objects.  Both are sufficiently
luminous ($\sim 0.8$ and $\sim 0.4$~mag brighter than the RGB tip) to
exclude the possibility of image blending.  However, both are also
located on the very outskirts of the the galaxy, where the likelihood
of contamination is greatest.  Since the density of intracluster stars
(immediately surrounding the dwarf) with $26 < {\rm F814W} < 27$ is
36~arcmin$^{-2}$, we should expect our sample to contain $\sim
1$~contaminating source in this magnitude range. Thus, it is uncertain
whether these two bright objects are actual members of the galaxy.

Two of the three remaining stars with $-4.2 < M_{bol} < -4.0$ are
projected near the center of the galaxy, and are thus likely to be
associated with the dSph.  If we apply the age-$M_{bol}$ relation of
\citet{rej06} to these objects, then the inferred age of the stellar
population is more than 8-10~Gyr. This is similar to the ages of the
old populations seen in Local Group dSphs \citep{mateo98}.  We cannot
exclude the possibility that these objects may be blends due to the
more crowded environment; if this is indeed the case, this would
strengthen the conclusion that the galaxy contains no stars with 
ages less than $\sim 10$~Gyr.

\subsection{Surface Brightness Profile}

\begin{figure}
\plotone{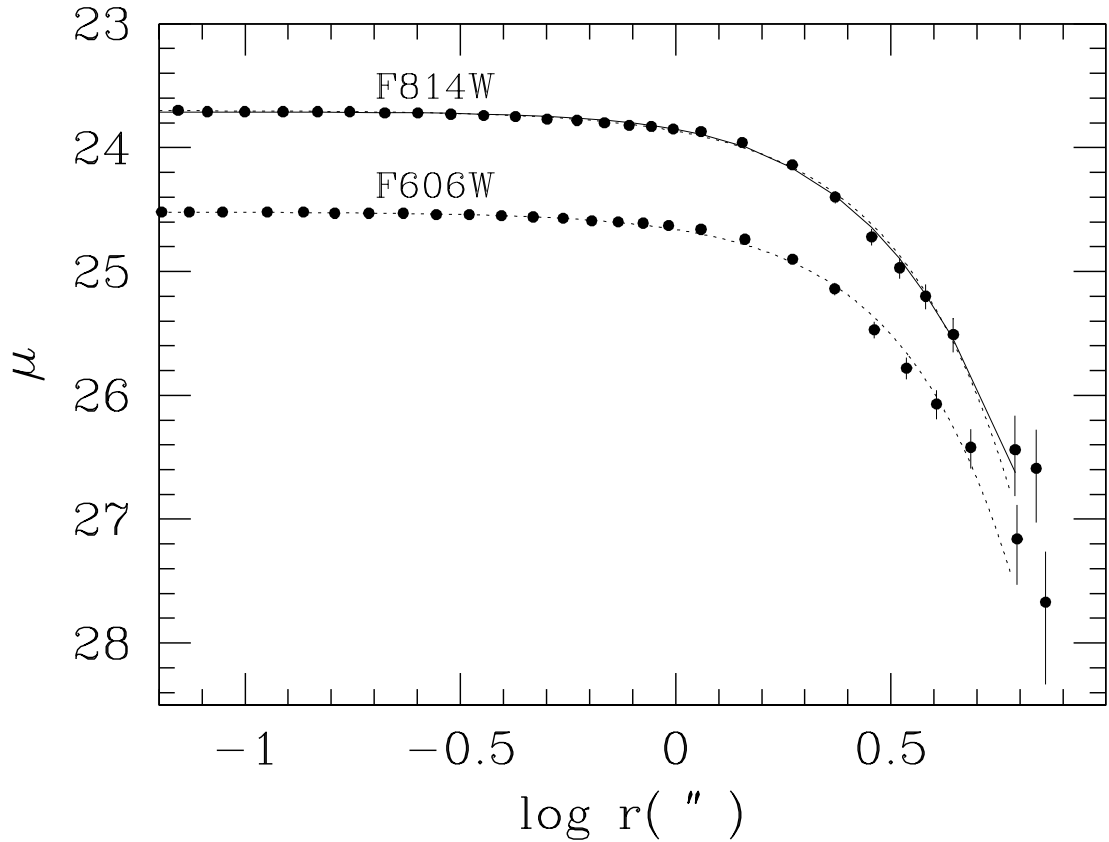}
\caption{F814W and F606W surface brightness profiles derived by fitting
elliptical contours to the smoothed images of the dSph galaxy.  The data
are plotted as a function of the geometric mean radius $r=\sqrt{ab}$.  The
best-fitting \citet{king62} model, with $r_c = 2 \farcs 6 \pm 0 \farcs 7$
and $r_t = 10\arcsec \pm 3\arcsec$ is shown as a solid line.  The 
\citet{king62} model fit is excellent, with $\chi^2/\nu = 0.40$.    The dashed 
lines denote the best fits to a \citet{sersic} profile; see text for details.
\label{profile}}
\end{figure}

In order to determine the basic structural parameters for our dSph, we
performed surface photometry on smoothed images of our frames.  To do
this, we began by increasing the per-pixel signal-to-noise of the
F606W and F814W images by re-binning the data $3 \times 3$.  We then
smoothed each frame with a circular Gaussian filter to produce images
with a $\sim 0 \farcs 6$ point-spread-function, and fit ellipses to
the dSph's contours using the {\tt ellipse} task in IRAF/STSDAS
\citep[][based on the algorithms of \citet{jedr87}]{busko96}.
To facilitate these fits, the center of the galaxy was kept fixed for
each contour, but the position angle and ellipticity were allowed to
vary to produce the best results.  In addition, to aid in the
measurement of this low surface brightness object, models were also
computed for three nearby background galaxies, whose light
contaminates the dSph's outer contours.  By subtracting these models
from the original image, we were able to improve the fits for the
target galaxy.  Finally, to compute the surface brightness of each
fitted contour, a sky background was determined using the median pixel
value derived within `empty' regions of our frame.
The resulting surface brightness profiles (out to a radius of $9 \farcs
3$ along the major axis) are plotted in Figure~\ref{profile} as a function
of the geometric mean radius ($r=\sqrt{ab}$).

Figure~\ref{profile} confirms that our galaxy is indeed a dwarf
spheroidal.  The central surface brightness of the galaxy is typical
of Local Group dwarf spheroidals \citep{mateo98}, with $\mu_0 = 24.54 \pm 0.03$ in
F606W and $\mu_0 = 23.70 \pm 0.03$ in F814W.
Moreover, as the dotted line in the figure demonstrates, the profile
of the galaxy is well fit by a \citet{king62} model.  To derive this
line, we convolved a series of \citet{king62} profiles with the
smoothed PSF, and fit the F814W data using all measurements with
errors less than 0.3~mag.  This procedure yields an excellent fit
$\chi^2/\nu = 7.9/20 = 0.40$, with a core radius of $r_c = 2 \farcs 6
\pm 0\farcs 7$ and a tidal radius of $r_t = 10\arcsec \pm 3\arcsec$
where the uncertainties are derived via Monte Carlo simulations
\citep[see][]{VICS2}.  At the distance of the dwarf, these measurements
correspond to $r_c = 220 \pm 80$~pc and $r_t = 850 \pm 250$~pc,
where the errors include the uncertainties in both the fit and the distance.

We have also fit our surface brightness profiles with the commonly used
\citet{sersic} profile \citep[see also][]{gd05}, \ie\ $\Sigma =
\Sigma_0 e^{{-(r/r_0)}^{1/n}}$.    Least squares fits to the F814W and 
F606W data yield shape parameters very close to the $n \sim 0.5$ value
expected for isothermal distributions: for the F814W image, $\Sigma_0
= 23.70 \pm 0.03$~mag~arcsec$^{-2}$, $r_0 = 3\farcs 16 \pm 0\farcs 16$
and $n = 0.61 \pm 0.12$, while the F606W profile has $\Sigma_0 = 24.52
\pm 0.03$~mag~arcsec$^{-2}$, $r_0 = 3\farcs 34 \pm 0\farcs 06$ and $n
= 0.59 \pm 0.12$.  Both fits (also plotted in Figure~\ref{profile})
are only slightly worse than those for the King model ($\chi^2/\nu =
0.59$ and $0.63$ for the F814W and F606W fits, respectively), but are
still a very good match to the data.

We can also use the data of Figure~\ref{profile} to compute the
integrated magnitude of our galaxy.   If we sum the flux contained
within each elliptical isophote, then the total magnitude of the dwarf
spheroidal is F606W$_{tot} = 20.56 \pm 0.05$ and F814W$_{tot} = 19.86
\pm 0.05$.  These values are consistent with the F606W $= 20.46$ and
F814W $=19.74$ magnitudes obtained from simple photometry using an
$11\arcsec$ circular aperture centered on the galaxy.  To convert
these magnitudes to $V$ and $I$, we used the color transformations of
\citet{sir05}, and to obtain absolute magnitudes, we applied the
distance and reddening values given in Section 3.1; no additional
corrections to our integrated magnitudes were required as our profiles
extend to very faint surface brightnesses.  The data imply that the
dSph galaxy has a total absolute magnitude of $M_V =-10.6 \pm 0.2$
(and $M_I=-11.4\pm 0.2$), where the error again includes both the
uncertainties of the photometry and the distance.

\subsection{Comparison with Local Group dSph Galaxies}

Detailed observations in the Local Group demonstrate that dSph galaxies 
possess a variety of star-formation histories and chemical abundances 
\citep{gw94, mateo98}.  This has been attributed to the effect of
environment on their formation and evolution.  If so, then
the properties of such galaxies in dense clusters might be distinctly
different from those of dSphs seen locally.   Unfortunately, although a
few dwarf galaxies have been studied in other nearby groups, such as the 
M81 system \citep{cald98,kks05, dacosta05}, very little is known about dSphs
in rich clusters.

\begin{figure}
\plotone{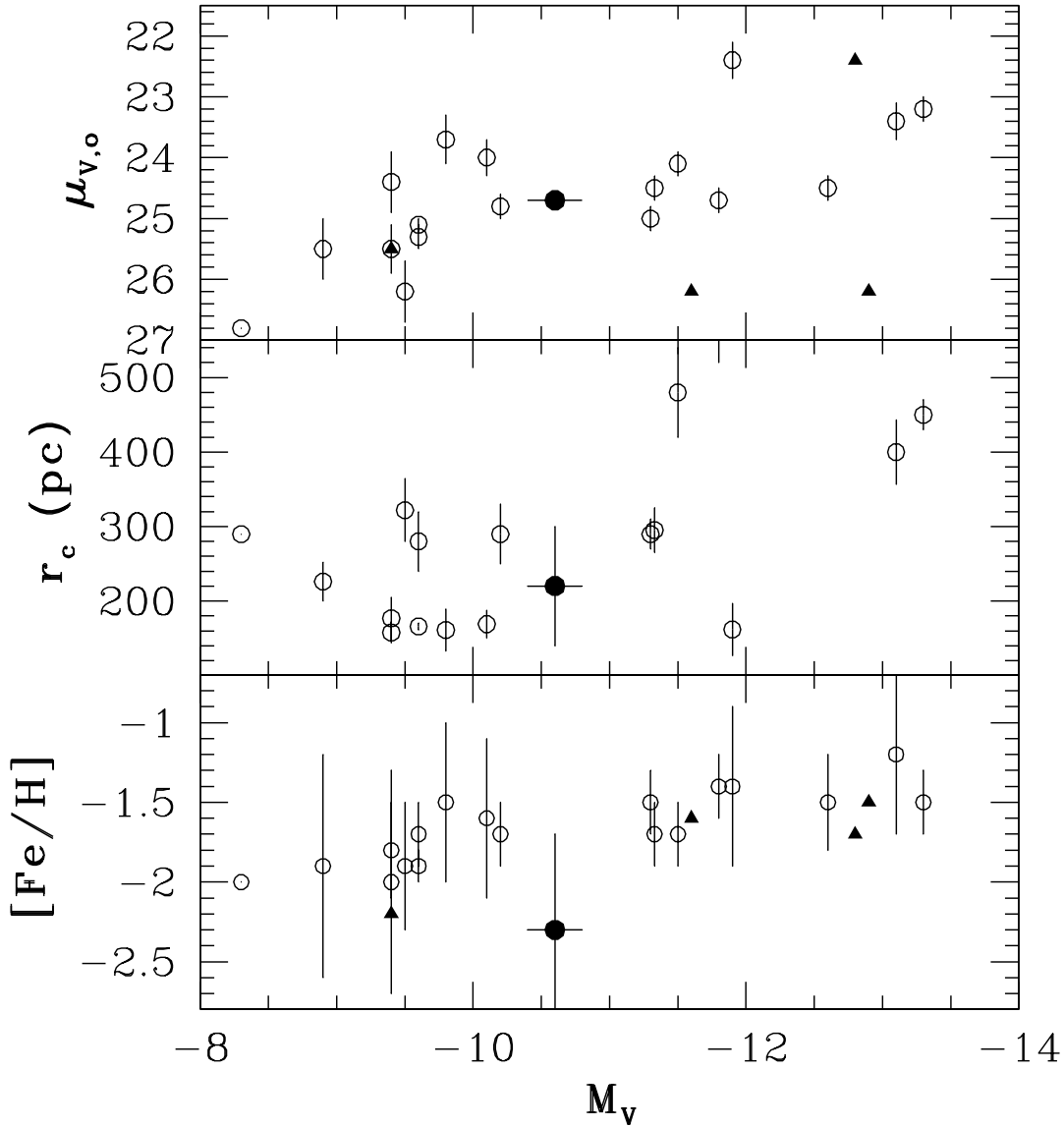}
\caption{A comparison of the structural and chemical 
properties of the Virgo dSph (filled circle) with measurements for the dwarf spheroidals in the Local 
and M81 Groups (open circles).  The top panel shows central surface brightness, the 
middle panel core radius, and the bottom panel metallicity, with the error 
bars representing the metallicity dispersion.  The filled triangles denote measurements of 
faint dE/dSph galaxies in the Virgo Cluster from \citet{cald06}.    The Local Group data come
from the compilations of \citet{ih95}, \citet{ggh03} and
\citet{mcc06}, with additional information for individual dwarfs from
\citet{sav96}, \citet{palma03}, and \citet{har05}.  The M81 dwarf data come
from \citet{cald98}.  Note that the properties of the Virgo dwarf are
completely consistent with those of dwarfs in these low-density systems.
\label{comparison}}
\end{figure}

Figure~\ref{comparison} compares the observed central surface
brightness, core radius, and metallicity of our Virgo dSph with the
corresponding properties of dSph galaxies in the Local and M81 Groups.
The most striking aspect of the figure is how normal our dSph appears.
The galaxy's structural parameters are similar to those of the Leo~II
and Ursa Minor dwarfs, \ie\ typical of objects present in the local
neighborhood.  This result is consistent with the studies of
\citet{dur97}, \citet{hmi03}, and \citet{cald06}, who also found no significant structural differences between 
the dwarfs of the Local Group and those of Virgo and Fornax.  For
comparison, the Virgo results from \citet{cald06} are also plotted in
Figure~\ref{comparison}.

We reach a similar conclusion from the metallicity measurement of our dSph 
galaxy.  Dwarf elliptical and spheroidal galaxies in the Local Group obey a
well-established luminosity-metallicity relation \citep[\eg][]{smith85,
cald92, cald98, cote00, cald06}, where the more luminous galaxies are
(on average) more metal-rich.   This result, which is usually attributed
to increased mass loss in systems with small potential wells
\citep[\eg][]{ds86}, predicts the metallicity of our dSph galaxy fairly
accurately.  However, galaxies in clusters are expected to lose much of
their mass to intracluster space via tidal encounters with other galaxies
and with the cluster as a whole \citep[\eg][]{moore96, moore98, bekki01}. If 
our dSph had been exposed to these forces for any length of time, we would 
expect its metallicity to be more in line with that of a higher-luminosity 
object.  This does not appear the case:  if anything, our galaxy's metallicity 
lies {\it below\/} the mean luminosity-metallicity relation.  This suggests
that most of the galaxy's original stars are still bound to the system, or 
perhaps the luminosity-metallicity relation differs in the cluster environment.

\begin{figure}
\plotone{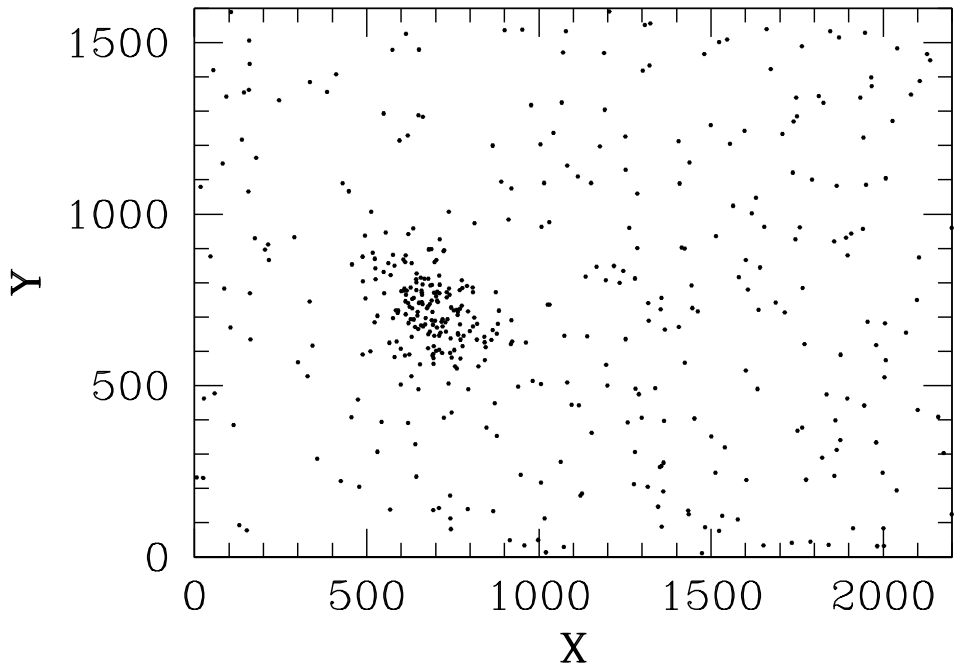}
\caption{Spatial distribution (in pixels) for those stellar objects 
near our dwarf galaxy with the colors and magnitudes of metal-poor
red giant stars (\ie\ $0.4 < {\rm F606W}-{\rm F814W} < 1.3$ and
$26.8 < {\rm F814W} < 28.0$).  Each pixels represents $0\farcs 03$,
or $\sim 2.2$~pc at the distance of Virgo.  There is no obvious evidence
for the tidal shredding of the dwarf galaxy.
\label{xyplot}}
\end{figure}

The structure of the galaxy appears to support this conclusion.  Tidal
features are often detected in dSph galaxies via photometric
deviations from a \citet{king62} profile \citep[\eg][]{ih95, palma03,
maj05}.  As Figure~\ref{profile} demonstrates, our Virgo dwarf is
well-fit by a King model all the way down to $\mu_V \sim 27.5$ ($\mu_I
\sim 26.5$), where the galaxy's surface brightness merges into that of
the intracluster background \citep{VICS1}.  Moreover, a plot of
($x,y$) positions using only those stars with colors and magnitudes
similar to those of the galaxy's RGB objects (\ie\ $0.4 < {\rm
F606W}-{\rm F814W} < 1.3$ and $26.8 < {\rm F814W} < 28.0$), reveals no
obvious evidence for tidal distortion (Figure~\ref{xyplot}).  We note
that our observations are not deep enough to see any low-level tidal
disruption of this dSph galaxy: the large streams seen in the Coma and
Centaurus Clusters are only visible at surface brightnesses below
$\mu_V \sim 26$ \citep{tm98, gw98, cal00}, and the tails associated
with the Local Group galaxies Carina and Ursa Minor are at $\mu_V
\gtrsim 29$ \citep{palma03, maj05}.  Thus we cannot unambiguously
state whether the dSph galaxy has undergone any tidal stripping.
Nevertheless, the metallicity arguments above suggest that the
observed dSph is not the low-luminosity remains of a much more massive
system (\eg\ as expected in galaxy harassment scenarios).

Tidal threshing is not the only process that could affect dwarf
galaxies in the cluster environment.  In the Local Group, ram pressure
stripping may be a prime factor in the evolution of dSph galaxies
\citep{lf83, vdb94, ggh03,marco03}.  Indeed, the N-body simulations of
\citet{mayer06} suggest that both tides and ram pressure are needed to
explain the properties of Local Group objects.  If true, then the
effects of intracluster gas should be magnified in the denser regions
of the Virgo Cluster core.  If low-luminosity dwarfs can barely hold
onto their gas (and enrich themselves) in the Local Group, then Virgo
dSphs should easily be stripped by the Virgo intracluster medium.  The
result would be a dSph galaxy much like those observed in the Local
Group (as suggested by the structural similarities), but with less
chemical enrichment, and a very narrow red giant branch.

Our dSph does show some evidence for enhanced pressure stripping: as
Figure~\ref{comparison} shows, the galaxy's mean metallicity is
slightly less than that predicted from dSph observations in the Local
and M81 Groups.  However, our RGB photometry is not precise enough to
test for self-enrichment.  Local Group dwarfs typically have
metallicity dispersions of $\gtrsim 0.3$~dex \citep{mateo98}.  Our
data, though consistent with a zero abundance spread, is best fit with
an RGB color dispersion of $\sigma_{{\rm F606W}-{\rm F814W}} = 0.05$
($\sim 0.4$~dex) and still admits a metallicity spread as great as
$\sigma_{{\rm F606W}-{\rm F814W}} = 0.09$ ($\sim 0.6$~dex).  Thus,
while the data are suggestive, we cannot say for certain that the
Virgo environment has had any additional effect on the dSph's chemical
properties.  Deeper photometric studies of other Virgo dSph galaxies
would be very useful in ascertaining any differences between the
chemical evolution of low-luminosity dwarf galaxies in clusters and
those in the group environment.

\section{Discussion}

From the analyses in the previous section, it is clear that the dSph
galaxy in our field lies in the Virgo cluster and appears relatively
undisturbed by its surroundings.  Its stellar population is very
metal-poor ([M/H] $\sim -2.3$) and old, with no indication of stars
younger than $\sim 10$ Gyr.  In this sense it is similar to the purely
old Local Group dSph galaxies of Ursa Minor and Draco.  It is also
similar to the dwarf galaxies recently observed by \citet{cald06} in
the Virgo Cluster core -- their central surface brightnesses and
metallicities are also included in Figure~\ref{comparison}.  They,
too, obey the luminosity-metallicity relation for small galaxies and
appear undisturbed by the cluster environment.

The appearance of such `pristine' old, low-luminosity dwarf galaxies
within the Virgo Cluster begs a question -- how do these objects last
so long in the cluster?  Small galaxies should be easy prey for the
tidal forces associated with the cluster environment \citep{moore98},
and, indeed, the velocity distribution of Virgo dE galaxies supports
the idea that at least some fraction of the cluster's dwarfs are the
stripped remains of larger, late-type systems \citep[\eg][]{con01}.
Furthermore, it is plausible that many of Virgo's metal-poor
intracluster stars \citep{VICS1} originated in such galaxies.  Nevertheless, 
it seems clear that some dwarfs remain intact; perhaps these 
are dwarfs that have recently fallen into the cluster, or have rather 
high M/L ratios.

A weak constraint on the history of our dSph galaxy can be obtained from
its observed tidal radius.   From \citet{king62}, the limiting radius  
of the galaxy should be
\begin{equation}
r_t = R_p \left( {M_g \over M_C (3 + \epsilon)} \right)^{1/3}
\end{equation}
where $M_g$ is the mass of the galaxy, $M_C$, the mass of the cluster,
$R_P$, the galaxy's pericluster radius, and $\epsilon$, the eccentricity
of the orbit.  The mass of the galaxy can be obtained by analogy with
Local Group objects:  according to \citep{mateo98}, dwarf galaxies with
$M_V \sim -10.6$ have mass-to-light ratios of $M/L_V \sim 10$, implying
$M_g \sim 1.5 \times 10^7 M_{\odot}$.  The Virgo cluster mass is
similarly obtainable from the x-ray observations of \citet{nb95}:  
their data give a core radius of 56~kpc, a mass per unit length of
$12 \times 10^{10} M_{\odot}$~kpc$^{-1}$, and a total mass within
180~kpc (the projected cluster distance of the galaxy) of
$2 \times 10^{13} M_{\odot}$.  A value of $\epsilon \sim 0.5$ then
excludes any orbit which extends into the inner
$\sim 80$~kpc of the cluster.

Of course, there are any number of ways to satisfy this condition.
For instance, the dSph may just now be falling into Virgo for the
first time.  Cepheids, the planetary nebula luminosity function, the
surface brightness fluctuation method, and the tip of the red giant
branch, all place the core of Virgo at a distance of between 14.5 and
17~Mpc \citep{freedman01, jacoby90, tonry01, jensen03, jbb04, cald06}.
Our dSph distance estimate of $17.6 \pm 1.4$~Mpc is consistent with
this range of values, but it also admits the possibility that the
galaxy is slightly behind the main body of the cluster.  If so, then
the dSph's location may associate it with the infalling M86 system,
which is $\sim 3$~Mpc beyond M87 \citep{bohringer94, jacoby90,
tonry01, jbb04}.  If this scenario is correct, then our dwarf galaxy
formed long ago in a smaller cluster/group, and the shredding of the galaxy 
is only now about to begin.

If the dwarf {\it did\/} originate and live much of its life in the cluster
environment, then it would have to have a high mass to withstand
the tidal forces of Virgo's densest regions.   This would require a
mass-to-light ratio that is much larger than the `normal'
value of $M/L_V \sim 10$ observed in the Local Group.  CDM models do
suggest that dwarf galaxies can form out of dark matter halos much more
massive than dynamical measurements suggest (\ie\ $M \sim 10^{9 - 10}
M_{\odot}$ \citep[\eg][]{stoehr02,hayashi03,kravtsov04}, and such objects would
be able to resist tidal disruption.  However, galaxies with large
dark matter halos would not easily be stripped of their ISM by ram
pressure or galactic winds.  \citet{mb00} point out that while low
mass ($M \sim 10^{6 - 7} M_{\odot}$) objects will rapidly lose their
gas to the intracluster medium, more massive systems are likely to hold
on to at least some of their interstellar medium.  As a result, 
their stellar populations should be younger and more metal-rich 
than their lower-mass counterparts.  While it is beyond the scope of
this paper to investigate the precise masses and conditions under
which early gas removal is expected, the old, metal-poor population of our
dwarf spheroidal argues against the presence of a very massive dark halo.   The 
existence of such a massive halo could, however, be tested via high-resolution 
spectroscopic observations with the next generation of large telescopes.

Unfortunately, our lack of knowledge about the dwarf's current
location and orbital characteristics precludes our placing a strong
constraint on its origin.  While we have no data on the presence
(or absence) of H~I gas in the galaxy, that gas has certainly not 
created many new stars in the past $\sim 8$~Gyr.  Thus, if gas were
detected, it would not affect our basic conclusions.

\section{Summary}

We report the discovery of a dwarf spheroidal galaxy (with
$M_V=-10.6\pm 0.2$) on deep \textsl{ACS} F606W and F814W images of an
intracluster field in the Virgo Cluster.  The distance to the galaxy
is $17.6\pm 1.4$ Mpc based on the location of its TRGB; this value is
consistent with a location close to the core of the Virgo cluster, although
we cannot rule out the possibility that this object is part of the M86
subcluster falling into Virgo from behind.  The galaxy is clearly
resolved into stars, and our observations extend more than a magnitude
down its red giant branch.

We find that the galaxy is composed entirely of old ($> 8-10$~Gyr), 
and is very metal-poor, with [M/H]$ = -2.3 \pm 0.3$.  This metallicity
is consistent with that expected for a galaxy of its luminosity; if
anything, the galaxy lies slightly below the mean luminosity-metallicity
relation.  Thus, the object is not the remains of a larger galaxy
that has been tidally stripped or harassed.  Moreover,
the dSph's structural properties are similar to those derived for
dwarf galaxies in the Local and M81 groups, suggesting that many of
the same physical processes that govern the formation and evolution of
local dSphs also apply to this Virgo Cluster object.  Based on 
this similarity, we suggest that our dSph galaxy is likely on its
initial infall into the center of the cluster.

\acknowledgments

The authors would like to thank Tom Brown for assistance regarding
dither patterns for the \textsl{ACS} observations, and Chris Palma for
reading an earlier version of this paper.  Support for this work was
provided by NASA grant GO-10131 from the Space Telescope Science
Institute and by NASA through grant No.\ NAG5-9377.  JJF was
supported by the NSF through grant AST-0302030, and TvH was supported
by NASA under grant No.\ NAG5-13070 issued through the Office of Space
Science.

{\it Facilities:} \facility{HST (ACS)}

\clearpage

\clearpage

\end{document}